

\vsize 8.7in

\def\doublespace{\baselineskip 22.76 pt}

\hsize 6.5 true in
\hoffset=0. true in
\voffset=0. true in
\def\mathnew{\mathsurround=0pt}
\def\simov#1#2{\lower .5pt\vbox{\baselineskip0pt \lineskip-.5pt
\ialign{$\mathnew#1\hfil##\hfil$\crcr#2\crcr\sim\crcr}}}
\def\simgreat{\mathrel{\mathpalette\simov >}}
\def\simless{\mathrel{\mathpalette\simov <}}
\font\twelverm=cmr10 scaled 1200
\font\ninerm=cmr7 scaled 1200
\font\sevenrm=cmr5 scaled 1200
\font\twelvei=cmmi10 scaled 1200
\font\ninei=cmmi7 scaled 1200
\font\seveni=cmmi5 scaled 1200
\font\twelvesy=cmsy10 scaled 1200
\font\ninesy=cmsy7 scaled 1200
\font\sevensy=cmsy5 scaled 1200
\font\twelveex=cmex10 scaled 1200
\font\twelvebf=cmbx10 scaled 1200
\font\ninebf=cmbx7 scaled 1200
\font\sevenbf=cmbx5 scaled 1200
\font\twelveit=cmti10 scaled 1200
\font\twelvesl=cmsl10 scaled 1200
\font\twelvett=cmtt10 scaled 1200

\def\batsefour{BATSE-Ch4}
\def\batsefifteen{BATSE-15$\sigma$}
\def\vvm{$V/V_{max}$}
\def\ffm{$(F_{min}/F_{max})^{3/2}$}

\def\vvmavg{$\langle V/V_{max}\rangle$}
\def\ffmavg{$\langle (F_{min}/F_{max})^{3/2}\rangle$}
\def\fak{$\langle (F_{min}/F_{max})^{3/2}_K\rangle$}
\def\fk{$(F_{min}/F_{max})^{3/2}_K$}

\def\qo{q$_\circ$}

\def\qomath{{\rm q}_\circ}
\def\Pacz{Paczy\'nski}

\skewchar\twelvei='177 \skewchar\ninei='177 \skewchar\seveni='177
\skewchar\twelvesy='60 \skewchar\ninesy='60 \skewchar\sevensy='60
\def\twelvepoint{\def\rm{\fam0 \twelverm}
  \textfont0=\twelverm \scriptfont0=\ninerm \scriptscriptfont0=\sevenrm
  \rm
  \textfont1=\twelvei \scriptfont1=\ninei \scriptscriptfont1=\seveni
  \def\mit{\fam1 } \def\oldstyle{\fam1 \twelvei}
  \textfont2=\twelvesy \scriptfont2=\ninesy \scriptscriptfont2=\sevensy
  \def\cal{\fam2 }
  \textfont3=\twelveex \scriptfont3=\twelveex
  \scriptscriptfont3=\twelveex
  \textfont\itfam=\twelveit \def\it{\fam\itfam\twelveit}
  \textfont\slfam=\twelvesl \def\sl{\fam\slfam\twelvesl}
  \textfont\bffam=\twelvebf \scriptfont\bffam=\ninebf
    \scriptscriptfont\bffam=\sevenbf \def\bf{\fam\bffam\twelvebf}
  \textfont\ttfam=\twelvett \def\tt{\fam\ttfam\twelvett}
  }
\def\folio{\ifnum\pageno=1\nopagenumbers\else\number\pageno\fi}

\font\twelvess=cmss10 scaled 1200
\twelvepoint
\doublespace
\def\ref{\par\noindent\hangindent=2pc \hangafter=1 }

\def\apj{{\it Ap.~J.}}
\def\apjl{{\it Ap.~J. (Letters)}}

\def\nature{{\it Nature}}

\null
\centerline{Submitted to the Editor of the Astrophysical Journal
{\sl Letters} on May 25, 1993}
\centerline{Revised July 8, 1993; Accepted July 23, 1993}
\null\vskip 0.85 true in
\centerline{\bf Reconciliation Of The Disparate Gamma-Ray Burst}
\vskip 0.1in
\centerline{\bf Catalogs In The Context Of A Cosmological Source
Distribution}
\vskip 1.0 true in
\centerline{{\bf Peter Tamblyn}\footnote
{\hbox{$\null^1$}}{ARCS Foundation Scholar.}{\bf \ and Fulvio
Melia}\footnote
{\hbox{$\null^2$}}{Presidential Young Investigator.}}
\vskip 0.05in
\centerline{\sl Steward Observatory, University of Arizona, Tucson AZ
85721}
\vfill\eject
\null
\centerline{\bf Abstract}
\bigskip
{\twelvess
It is well known that Gamma-ray Burst spectra often display a break at
energies
$\simless 400$ keV, with some exceptions extending to several
MeV.\ \ Modeling
of a cosmological source population is thus non-trivial when comparing
the
catalogs from instruments with different energy windows since this
spectral
structure is redshifted across the trigger channels at varying levels
of
sensitivity.  We here include this important effect in an attempt to
reconcile
all the available data sets and show that a model in which bursts have
a ``standard''
spectral break at $300$ keV and occur in a population uniformly
distributed in a
\qo $=1/2$ universe with no evolution can account very well for the
combined set of
observations.  We show that the source population cannot be truncated
at a minimum
redshift $z_{min}$ beyond $\sim 0.1$, and suggest that a simple
follow-on
instrument to BATSE, with the same trigger window, no directionality
and 18 times
better sensitivity might be able to distinguish between a \qo$=0.1$ and
a \qo$=0.5$ universe in $3$ years of full sky coverage, provided the
source
population has no luminosity evolution.
}
\bigskip\noindent {\it Subject headings}:  cosmology: observations --
cosmology: theory -- galaxies: clustering --
galaxies: evolution -- gamma rays: bursts -- pulsars

\vfil\eject
\centerline{\bf 1. Introduction}
\medskip
Observations with the Burst and Transient Source Experiment (BATSE) on
the Compton
Gamma Ray Observatory (CGRO) reveal an isotropic distribution of
Gamma-ray Bursts
(GRBs) which nonetheless appears to be radially truncated in the
context of a
Euclidean geometry (Meegan et al. 1992).  This seems to rule out nearby
(i.e.
Galactic disk) single population models, and has led to renewed
speculation that GRBs
originate at large redshifts (Usov \& Chibisov 1975).  Recently,
the cosmological hypothesis has been revisited by, e.g.
Mao \& Paczy\'nski (1992), Piran (1992), and Dermer (1992), who
considered the relation
between the geometrical distribution of GRB sources and the statistics
of the
observed bursts under the simplifying assumption that all the bursts
are
standard
candles with identical power-law spectra.  In particular,
Mao \& Paczy\'nski (1992) showed that the simplest cosmological
model (i.e., a flat universe with $\Omega=1$ and a cosmological
constant
$\Lambda=0$) corresponds reasonably well with the intensity
distribution
of weak and strong bursts observed with BATSE and PVO (Epstein \&
Hurley 1988).

How these bursts might be produced is not at all clear, though several
plausible physical
scenarios continue to evolve.  Non-catastrophic processes require
focusing of the emitted
energy, such as would occur in sheared Alfv\'en wave dissipation near
the polar
cap of highly-magnetized neutron stars which produce streams of
relativistic particles
that are beamed by the underlying magnetospheric structure (Melia \&
Fatuzzo 1992).
In this case, the GRB spectrum results from the Compton upscattering of
the
corresponding radio pulsar radiation, which is often characterized by a
spectral turnover
at $\sim 1$ GHz.  This feature should therefore manifest itself as a
break at gamma-ray
energies ($\epsilon_b\sim$ several hundred keV) if the energized
particles have a Lorentz factor
$\sim 10^{5-6}$.

What {\it is} clear from the sample of bursts observed with BATSE's
Spectroscopy Detectors
is that GRB spectra are typically well described by broken power laws
with a turnover
energy $E_\circ$ ranging from below $100$ keV to more than an MeV, but
peaking
under $200$ keV with only a small fraction of the spectra breaking
above $400$ keV
(Band et al. 1993).  The impact of this spectral structure on the
modeling of
a cosmological population can be substantial, particularly when
comparing the
catalogs from instruments with different energy windows, since the
break feature
necessarily redshifts into and out of the observed energy range at
varying
levels of sensitivity.  We here consider this important effect and its
implications
on the cosmological hypothesis, and discuss the limits one may thereby
reasonably
place on the evolution of the comoving source density and luminosity
function,
and on the redshift range sampled by each detector.
An earlier treatment of the impact of redshift on the observed
spectra of individual bursts was presented by Paczy\'nski (1992).

\medskip
\centerline{\bf 2. The Data Samples And Model Source Distributions}
\medskip

Due to the short time that BATSE has been collecting data, its catalog
contains only a few of
the important rare, strong bursts and must therefore be supplemented
with
bursts recorded by PVO, SMM, KONUS, SIGNE, and APEX.  The use of data
from multiple instruments (with uncertain relative sensitivities)
introduces considerable complication due to varied burst identification
criteria.
In addition, intercomparison requires correction of the observed
rates to full sky, live time equivalent rates.  The \vvmavg\ test
(Schmidt
et al. 1988)
allows comparison between disparate instruments
despite uncertain relative sensitivities.  \vvm\
corresponds directly to a ratio of sampling volumes
only in Euclidean space, and
throughout the following discussion we will
refer to this statistic as \ffm\ whether it is determined from peak
photon
or energy fluxes or fluences, except where the distinction is
important.
Each of the selected data sets (summarized in Table I) has a large
sample of
bursts, published \ffmavg\ values and corrected detection rates ${\cal
R}$.  The errors
in ${\cal R}$ were estimated by scaling the square root of the burst
sample size from which ${\cal R}$ was determined and represent
lower limits to the uncertainties.

The BATSE data currently available in the
public domain are sufficiently detailed to allow the definition of two
useful
subsamples.  In constructing each subsample, we have omitted bursts
with
incomplete information or with an overwrite flag.

The first, \batsefifteen, simulates a less sensitive detector by
setting
$C_{lim}$ to 15 times the background uncertainty instead of the
5.5$\sigma$\
used for BATSE triggered sample (Meegan et al. 1992) and still contains
more
than 100 bursts.
The second subsample, \batsefour, consists of those bursts
with a fluence in the fourth LAD spectral channel ($300 - \approx 1000$
keV; Fishman et al. 1989),
greater than $5.5$\ times the error in this fluence as listed in the
burst catalog.
Sub-sample detection
rates were scaled from the BATSE rate based on the
fraction of bursts which passed the secondary criterion.

Our analysis is based on models of the source distribution at
cosmological
distances similar to those discussed by, e.g., Mao \& \Pacz\ (1992),
but with the following important
distinctions.  First, while we only consider standard cosmologies with
$\Lambda = 0$, we do permit \qo\ to vary.
Second and more significantly, we approximate the intrinsic burst
spectrum
as a broken (rather than a single) power law
$$
\nu L_\nu d\nu = \cases{A\nu^{\alpha_1}d\nu &$\nu \leq \nu_b$\cr
\null&\null\cr
A \nu_b^{\alpha_1 - \alpha_2} \nu^{\alpha_2} d\nu &$\nu \geq
\nu_b$}\;,\eqno(1)
$$
and integrate over a fixed
detection bandpass $\nu_1 (1 + z)$ to $\nu_2 (1+z)$.
In their study of 54 bursts observed with BATSE, Band et al.\ (1993)
found that the
spectral index we label $\alpha_1$\ ranges from 0.5 to 2.7 with the
majority of
the events at $\approx 1$, and that $\alpha_2$\ ranges from more than 0
to less than $-3$\
with most of the bursts at just under $0$ (see also Schaefer 1992).  We
assume the fiducial values of
1 and $-1$\  for these indices, respectively, for which Band et al.'s
(1993) $E_\circ$\
(see \S 1 above) is then half the break energy $E_b$ in the observer's
frame.
The models we discuss assume an intrinsic break energy
$\epsilon_b\equiv h\nu_b = E_b (1+z) = 300$~keV.
\medskip
\centerline{\bf 3. Analysis And Discussion}
\medskip

For each distribution, we use an iterative technique to identify
BATSE's limiting redshift
$z_{max}$, such that the corresponding value of $F_{min}=F(z_{max})$
gives the correct
\ffmavg.  This specifies the value of $A$ relative to $F_{min}$, and
integration
over the volume up to $z_{max}$ fixes the density $n_0$ such that the
rate ${\cal R}$
matches that observed by BATSE.\ \ A model is generated by varying the
limiting flux, simulating
more or less sensitive detectors, and integrating \ffmavg\ and burst
rate out to the new limiting redshift.  Predictions of what other
experiments would measure
are made by integrating over the same source distribution (with
identical $n_0$ and $A$),
but with $F(z)$ integrated over a different
bandpass.  Figure 1 shows the observed data points listed in Table I
along with the
corresponding \ffmavg\ -- ${\cal R}$ curves for the simplest model with
a spectral
break (i.e., for a distribution of standard candle sources with a
single intrinsic spectrum broken
at $\epsilon_b=300$ keV and uniformly distributed in a $\qomath =
$\ 1/2
universe with no evolution).  It is important to emphasize that without
a spectral break,
all the curves would be degenerate and could not meet all of the
data points within the indicated $1\,\sigma$ error bars.

Though a single model source distribution is used for Figure 1, each
data point is associated
with a different curve because different instruments ``see'' different
parts of
the spectrum.  To clarify the presentation, we can instead
``K-correct'' the data to a standard passband
corresponding to the BATSE curve shown in Figure 2.  To do this,
we first construct a model as described above and estimate the limiting
redshift for a given
experiment so that enough sources are included to reproduce the
observed rate.
Then for each individual burst in each data set, its \ffm\ and the
corresponding instrument's
$F_{min}$ give the flux in the observed bandpass.  The redshift of that
source can be found
in the case of a standard candle model with weak or no luminosity
evolution because the flux is
a monotonically decreasing function of $z$.  The spectral model,
redshifted
appropriately, can be used to determine what flux $F_{max,K}$ would be
measured
for that same burst in BATSE's bandpass.  The ratio of $F_{max,K}$ over
$F_{min,K}$
then yields the K-corrected \fk.
Averaged over all the bursts in a sample, these give the corresponding
K-corrected \fak\ shown in Figure 2.\ \
For a single detector, the flux--distance relationship
is systematically distorted (depending on the energy window) by the
redshift of the break across the
bandpass,i.e., more distant bursts
have observed fluxes progressively lower than would be expected from
the
luminosity distance alone.
Therefore, when the K-correction is invoked to directly
compare \ffmavg\ vs.\ ${\cal R}$ for
different instruments, the calculated \fak\ in this figure
is different from \ffmavg.
SIGNE is not included on this plot because the data for
individual bursts are not available.

If GRB locations correlate with luminous matter, their distribution
might be expected to appear clumped on the sky.  However, the angular
correlation function for the burst positions in two catalogs (Hartmann
\& Blumenthal 1989) does not show any clustering and indicates a
minimum
distance scale of $\simgreat 100$ Mpc (or a minimum redshift
$z_{min}\simgreat 0.05$, see Eq. 1)
for the {\it fainter} bursts.  This suggests that
the source population may be truncated with no members in the local
universe.  We therefore consider the implications of a truncated
population
by integrating from a non-zero minimum redshift $z_{min}$\ when
calculating the
predicted \fak\ -- ${\cal R}$ curves, shown in Figure 3 including the
K-corrected data points from Figure 2.  This figure clearly illustrates
that
a population truncated at a $z_{min}$\ $\simgreat 0.1$ is inconsistent
with the available data and a simple cosmological source
model.  The {\it brightest} bursts must originate
nearby if this model is valid.

\bigskip
\centerline{\bf 4. Conclusions}
\medskip
In this paper, we have attempted to reconcile the various GRB catalogs
with a single unifying
cosmological source distribution and have shown that a simple model in
which bursts have
a spectral break at $300$ keV and occur in a population uniformly
distributed in a
\qo $=1/2$ universe with no evolution can account very well for all the
observed characteristics.
The qualitative aspects of our results are not changed by the inclusion
of a model with a
distribution in $\epsilon_b$, as long as this distribution peaks below
$\sim 800$ keV, but
the fit improves as $\epsilon_b$ approaches $\sim 300$ keV.
Although the Ginga data set could not be included due to the absence of
precise information
concerning its dead time correction and field of view obscuration, we
note that the estimated
improvement of a factor of 2-3 in its sensitivity over KONUS and its
observed value of $0.35\pm0.035$ in \ffmavg\ (Ogasaka et al. 1991)
would place it between
the BATSE-15$\sigma$ and BATSE data points in Figure 2, and would
therefore be fully consistent
with the model.  We note that without consideration of the
spectral break, the multiplicative probability that all of the data
points
are consistent with the simple cosmological curve is $\sim 0.01\%$.
By comparison, the likelihood for consistency with a single
model using the break improves to the significantly greater value of
$\sim 3\%$.\ \ Even excluding the SMM value, these probabilities
are 0.3\% and 6\%, respectively.  These would presumably
improve if the other differences between these experiments (e.g.
threshold
effects and trigger time scales) and the intrinsic source properties
(e.g. luminosity function) were incorporated into the analysis.

An equally important result of our analysis is the estimation of a
maximum value for the
minimum redshift $z_{min}$ of the source population.  We have seen that
a distribution
truncated at $z_{min}\simgreat 0.1$ is inconsistent with the combined
body of data considered
here.  As such, the absence of
M31-like galaxies within the error boxes of known burst locations
(Schaefer 1992)
might argue against a dominant association of GRB sources with large
galaxies.
However, these objects could reside in LMC-scale (or smaller) galaxies,
which are known to be statistically
associated with their larger brethren.  The fields with the
tight burster error boxes should be reanalyzed to determine if there
is a statistical excess of large galaxies {\it near}\ the error boxes.
This correlation would suggest that the sources are associated with
small galaxies
rather than with massive galaxy cores.

It is also possible to limit the evolution of
GRB sources in the context of these models by requiring a $1\sigma$ fit
to all
the data sets previously described (Tamblyn \& Melia 1993).
Within the redshift of
$\approx 1$ sampled by BATSE, the differences introduced by reasonable
\qo\ variations (i.e., $0.1\simless$ \qo $\simless 0.9$) are dwarfed by
evolutionary uncertainties.
However, we remark on a potentially interesting future experiment
that may be able to distinguish at least between the cases \qo$\sim
0.1$ and \qo$\sim 0.5$,
provided only that there is no luminosity evolution in the source
population, though other
factors, such as number evolution, the break energy and $z_{min}$ may
vary.
Assuming BATSE survives long enough to detect $\approx 800$ bursts, so
that $n_0$ is known to
roughly $10\%$, a simple instrument with no directionality and roughly
$18$ times BATSE's
sensitivity in the same bandpass would detect a sufficient number of
events in
$3$ years of full sky coverage to differentiate between the measured
\ffmavg\ for
these two values of \qo\ at $\ge 3\,\sigma$.
Such an experiment might fall within the confines of a NASA SMEX
mission as an appropriate
follow-on to BATSE.

We are grateful to N. Bhat for helpful discussions and to S. Matz for
providing the SMM data.
The referee made several helpful comments.
This research was supported by NSF grant PHY 88-57218, the NASA
HE ADAP, and the ARCS Foundation.
\vfill\eject\null
{\baselineskip 13pt
\centerline{\bf References}
\medskip
\ref Atteia et al. 1991, \nature, {\bf 351}, 296.
\ref Band, D. 1993, in preparation.
\ref Band, D. et al. 1993, \apj, in press.
\ref Chuang, K.W. 1990, Ph.D. thesis, Univ. of California at
Riverside.
\ref Chuang, K.W. et al. 1992, \apj, {\bf 391}, 242.
\ref Dermer, C.D. 1992, {\it Phys. Rev. Lett.}, {\bf 68}, 1799.
\ref Epstein, R.I. \& Hurley, K. 1988, {\it Ap. Letters Comm.}, {\bf
27}, 229.
\ref Fenimore, E.E., Epstein, R.I., Ho, C. Klebesadel, R.W. \& Laros,
J. 1992, \nature, {\bf 357}, 140.
\ref Fishman, G.J. et al. 1989, in {\it Proceedings of the Gamma-Ray
Observatory
Science Workshop}, 2-39.
\ref Hartmann, D. \& Blumenthal, G.R. 1989, \apj, {\bf 342}, 521.
\ref Higdon, J.C. \& Schmidt 1990, \apj, {\bf 355}, 13.
\ref Mao, S. \& Paczy\'nski, B. 1992, \apjl, {\bf 388}, L45.
\ref Matz, S. et al. 1992, {\it Proc. Los Alamos
Workshop on Gamma Ray Bursts}, ed. C. Ho, R. Epstein, \& E. E. Fenimore
(Cambridge: Cambridge Univ. Press), p. 175.
\ref Matz, S. 1993, private communication.
\ref Meegan, C.A. et al. 1992, \nature, {\bf 355}, 143.
\ref Melia, F. \& Fatuzzo, M. 1992, \apjl, {\bf 398}, L85.
\ref Mitrofanov et al. 1992, {\it Proc. Los Alamos
Workshop on Gamma Ray Bursts}, ed. C. Ho, R. Epstein, \& E. E. Fenimore
(Cambridge: Cambridge Univ. Press), p. 203.
\ref Mitrofanov et al. 1991, {\it Sov. Astron.}, {\bf 35(4)}, p. 367.
\ref Ogasaka, Y., Murakami, T. \& Nishimura, J. 1991, \apjl, {\bf 383},
L61.
\ref Paczy\'nski, B. 1992, \nature, {\bf 355}, 521.
\ref Piran, T. 1992, \apjl, {\bf 389}, L45.
\ref Schaefer, B.E. 1992, \apjl, {\bf 393}, L51.
\ref Schaefer, B.E. 1992, {\it Proc. Los Alamos
Workshop on Gamma Ray Bursts}, ed. C. Ho, R. Epstein, \& E. E. Fenimore
(Cambridge: Cambridge Univ. Press).
\ref Schmidt, M., Hidgon, J.C., \& Hueter, G. 1988, \apjl, {\bf 329},
L85.
\ref Tamblyn, P. \& Melia, F. 1993, in preparation.
\ref Usov, V.V. \& Chibisov, G.V. 1975, {\it Astron. Zh.}, {\bf 52},
192.
\vfill\eject\null
\centerline{\bf Table I}
\vskip 0.2in
\centerline{Gamma-ray Burst Data Sets}
\vskip 0.2in\hrule\vskip 0.03in
\settabs\+\noindent&Experiment$^1$\quad&Bandpass (keV)\quad&Sample
Size\quad&
      ${\cal R}^{\dag}$ (yr$^{-1}$)\quad\quad&$\langle
      (F_{min}/F_{max})^{3/2}\rangle$\quad&Criteria\cr
\+&Experiment$$\quad&Bandpass (keV)\quad&Sample Size\quad&
      ${\cal R}^{\dag}$ (yr$^{-1}$)\quad&$\langle
      (F_{min}/F_{max})^{3/2}\rangle$\quad&$z_{max}$\cr
\vskip 0.03in\hrule\vskip 0.03in
\settabs\+\noindent&Experiment$^1$\quad&Bandpass (keV)\quad&Sample
Size\quad&
      ${\cal R}^{\dag}$ (yr$^{-1}$)\quad\quad&$\langle
      (F_{min}/F_{max})^{3/2}\rangle$\quad&Criteria\cr
\+&PVO$^1$&100 -- 2000&228 &$\,$\ 38$\pm$\ 3&0.46\ $\pm$0.02&0.2\cr
\+&SMM$^2$&350 -- 800&132 &$\,$\ 50$\pm$\ 5&0.40\ $\pm$0.025&0.3\cr
\+&KONUS$^3$&$\,$\ 50 -- 150&123 &130$\pm$12&0.45\ $\pm$0.03&0.4\cr
\+&SIGNE$^4$&$\,$\ 50 -- 400&169 &125$\pm$10&0.42\ $\pm$0.02&0.4\cr
\+&APEX$^5$&120 -- 700&58 &115$\pm$16&0.39\ $\pm$0.04&0.4\cr
\+&BATSE-Ch4$^6$&300 -- $\approx$1000&43
&168$\pm$26&0.35\ $\pm$0.044&0.4\cr
\+&BATSE-15$\sigma$$^6$&$\,$\ 50 -- 300&115
&430$\pm$43&0.384$\pm$0.029&0.7\cr
\+&BATSE$^7$&$\,$\ 50 -- 300&153; 271
&800$\pm$65&0.335$\pm$0.018&1.0\cr
\vskip 0.03in\hrule
\vskip 0.15in
\noindent\null$^{\dag}$Errors are based on Poisson statistics,
with $\Delta {\cal R}/{\cal R}=\sqrt{N}/N$, where $N$ is the
sample size\vskip 0pt
\noindent\null$^1$Chuang et al. (1992)\vskip 0pt
\noindent\null$^2$Matz et al. (1992);  Matz (1993)\vskip 0pt
\noindent\null$^3$Higdon \& Schmidt (1990); Matz et al. (1990)\vskip
0pt
\noindent\null$^4$Mitrofanov et al. (1991); Atteia et al. (1991)\vskip
0pt
\noindent\null$^5$Atteia et al. (1991); Mitrofanov et al. (1991);
Mitrofanov et al. (1992)\vskip 0pt
\noindent\null$^6$Fishman et al. (1989); this work\vskip 0pt
\noindent\null$^7$Fishman et al. (1989); Meegan et al. (1992); Band
(1993)\vskip 0pt
}
\vfill\eject\null
\centerline{\bf Figure Captions}
\vskip 0.1in\noindent{\bf Fig. 1.} -- Observed values of
\ffmavg\ versus the detectable burst rate ${\cal R}$
for each of the 8 samples considered here.  The relevant energy windows
for the various detectors are indicated
in the figure.  The corresponding theoretical curves are calculated for
the simplest model for BATSE
normalization for $n_0$, consisting of
a distribution of standard candle sources with a single intrinsic
spectrum broken at $\epsilon_b=300$ keV
and uniformly distributed in a \qo $=1/2$ universe with no evolution.
Each of the curves passes within
$1$ sigma of its associated data point.
\vskip 0.1in\noindent{\bf Fig. 2.} -- Same as Figure 1, except that now
only the curve corresponding to
BATSE's bandpass is shown.  All the data points are K-corrected to this
same energy window so that they too
correspond to the single curve shown here.  As was the case in Figure
1, all the data are within $1$ sigma
of the theoretical prediction.
\vskip 0.1in\noindent{\bf Fig. 3.} -- Same as Figure 2, except that the
theoretical curves are now calculated
under the assumption that the source population is locally truncated,
i.e., \fak\ is determined by integrating
from a non-zero minimum redshift $z_{min}$.  A population truncated at
a $z_{min}\simgreat 0.1$ is
inconsistent with the available data and a simple cosmology.
\vfill\eject\null
\pageno 15\hbox{\bf Authors' Addresses:}
\vskip 0.3 in
\settabs\+\noindent&{\bf Peter Tamblyn:}\quad&\cr
\+&{\bf Fulvio Melia \&}&\null\cr
\+&{\bf Peter Tamblyn:}&Steward Observatory, The University of Arizona,
Tucson AZ 85721\cr
\end